%% file: susyb.tex
\RequirePackage{lineno}
\documentclass[twocolumn,10pt,5p]{elsarticle}
\graphicspath{{figures/}}

\usepackage{atlasphysics} 

\usepackage{subfigure}
\usepackage{mathrsfs}
\usepackage{graphicx}
\usepackage{multirow}
\usepackage{rotating}
\usepackage{xspace}

\newcommand{\gl}   {\tilde{g}}
\newcommand{\sq}   {\tilde{q}}
\newcommand{\sqR}   {\tilde{q}^{}_{R}}
\newcommand{\sqL}   {\tilde{q}^{}_{L}}
\newcommand{\sqonetwo}   {\tilde{q}^{}_{1,2}}
\newcommand{\bone}   {\tilde{b}^{}_{1}}
\newcommand{\tone}   {\tilde{t}^{}_{1}}
\newcommand{\neut}  {\tilde{\chi}^{0}_{1}}  
\newcommand{\chipm}  {\tilde{\chi}^{\pm}_{1}} 
\def\dphimin{\ensuremath{\Delta\phi_{min}}}
\def\trmass{\ensuremath{m_\mathrm{T}}}
\def\bjets{$b$-jets\xspace}

\def\cjets{$c$-jets\xspace}

\def\ljets{light-jets\xspace}

\def\msugra{MSUGRA/CMSSM\xspace}
\def\btagging{$b$-tagging\xspace}

\def\meff{\ensuremath{m_{\mathrm{eff}}}\xspace}
\def\etmiss{\ensuremath{E_{\mathrm{T}}^{\mathrm{miss}}}\xspace} 
\renewcommand{\met} {\ensuremath{E_{\mathrm{T}}^{\mathrm{miss}}}\xspace}
\renewcommand{\pt} {\ensuremath{p_{\mathrm{T}}}\xspace}
\newcommand{\btheta}{\bf \theta}

\journal{Physics Letters B}
\newcommand{\susybtitle}{Search for supersymmetry in pp collisions at $\sqrt{s} = 7\TeV$ in final states with
              missing transverse momentum and b-jets}

\begin{document}
\begin{frontmatter}
  

\title{\susybtitle}

\author{The ATLAS Collaboration}


\begin{abstract}

Results are presented of a search for supersymmetric particles in events with large missing transverse
momentum and at least one heavy flavour jet candidate in $\sqrt{s} = 7$~TeV proton-proton collisions.
In a data sample corresponding to an integrated luminosity of 35~$\ipb$ recorded by the ATLAS experiment at the Large Hadron Collider, no significant 
excess is observed with respect to the prediction for Standard Model processes. 
For $R$-parity conserving models in which
sbottoms (stops) are the only squarks to appear in the gluino decay
cascade, gluino masses below 590~GeV (520~GeV) are excluded at the 95\% C.L.
The results are also interpreted in an \msugra supersymmetry breaking scenario with $\tan \beta=$40 and in an
SO(10) model framework.

\end{abstract}

\end{frontmatter}

\section{Introduction}
\input{intro_paper}

\section{The ATLAS Detector}
\input{detector}

\section{Simulated Event Samples}\label{sec:SimEvSamp}

\input{MCsample_paper}

\section{Data and Event Selection}
\input{data_evtsel}
\section{Standard Model Background Estimation}\label{sec:wbkg}
\input{bkg_paper}

\section{Systematic Uncertainties}
\input{syst_paper}

\section{Results}
\input{results}

\section{Conclusions}

\input{conclusion}

\input{Acknowledgement-09Mar11-collisions}

\bibliographystyle{atlasnote}
\bibliography{susyb}

\onecolumn
\clearpage

\begin{flushleft}
{\Large The ATLAS Collaboration}
\input{atlas_authlist}

\end{flushleft}
\end{document}

%% file: intro_paper.tex
Supersymmetry (SUSY)~\cite{Golfand:1971iw} is one of the most compelling theories 
to describe physics beyond the Standard Model (SM). It naturally solves the hierarchy 
problem and provides a possible candidate for dark matter. 
SUSY is a symmetry that relates fermionic and bosonic degrees of freedom, and postulates the existence of superpartners for the SM particles. 
Experimental data imply that supersymmetry is broken and that the 
superpartners are expected to be heavier than the SM partners. 
In the framework of a generic $R$-parity conserving minimal 
supersymmetric extension of the SM, the MSSM~\cite{Martin:1997ns}, SUSY particles are produced in pairs and 
the lightest supersymmetric particle (LSP) is stable. In a large variety of models, 
the LSP is the lightest neutralino, $\neut$, which is only weakly interacting. 

If supersymmetric particles exist at the TeV energy scale, the coloured superpartners
of quarks and gluons, the squarks ($\sq$) and gluinos ($\gl$), 
are expected to be copiously produced via the strong interaction at the 
Large Hadron Collider (LHC) \cite{ATLAS-CSC, CMS-TDR}.
Their decays via cascades ending with the LSP produce striking experimental signatures leading to 
final states containing  multi-jets, missing transverse momentum (its magnitude is referred to as 
$\etmiss$ in the following) -- resulting from the undetected
neutralino -- and possibly leptons.
First searches for the production of SUSY particles at the LHC have been 
published recently \cite{ATLAS-SUSY01,ATLAS-SUSY02, CMS-SUSY}. 

In the MSSM, the scalar partners of right-handed and left-handed quarks, $\sqR$ and $\sqL$, 
can mix to form two mass eigenstates. 
These mixing effects are proportional to the corresponding fermion masses 
and therefore become important for the third generation. In particular, large mixing 
can yield sbottom ($\bone$) and stop ($\tone$)  mass eigenstates which are 
significantly lighter than other squarks. Consequently, $\bone$ and $\tone$ could be produced 
with large cross sections at the LHC, either via 
direct pair production or, if kinematically allowed, through $\gl\gl$ production with 
subsequent $\gl \rightarrow \bone b$ or  $\gl \to \tone t$ decays.
Depending on the SUSY particle mass spectrum, the cascade decays of gluino-mediated and 
pair-produced sbottoms or stops result in complex final states consisting of $\etmiss$, several jets, among 
which $b$-quark jets ($b$-jets) are expected, and possibly leptons. 

In this letter, a search for final states involving $\etmiss$ and
$b$-quark jets is discussed. 
Results on searches for direct sbottom~\cite{sbottomCDF,sbottomD0}, stop~\cite{stopCDF,stopD0} 
and gluino mediated production~\cite{gluinosbottomCDF} have been previously reported by the 
Tevatron experiments, placing exclusion limits on the mass of these particles in several 
MSSM scenarios. 

The search described here is based on $pp$ collision data at a centre-of-mass energy 
of 7~TeV recorded by the ATLAS experiment at the LHC in 2010. The total data set corresponds   
to an integrated luminosity of 35~$\ipb$.
To enhance the sensitivity to different SUSY models, the search was performed using two  
mutually exclusive final states, characterised by the presence of leptons.
They are referred to as zero-lepton and one-lepton analyses in the following. 

In the zero-lepton analysis, events are required to contain energetic 
jets, of which one must be identified as a $b$-jet, large $\etmiss$ and no isolated 
leptons ($e$ or $\mu$). The zero-lepton analysis is employed to search for gluinos and sbottoms  
in MSSM scenarios where the $\bone$ is the lightest squark, all other 
squarks are 
heavier than the gluino, and $m^{}_{\gl}>m^{}_{\bone}>m^{}_{\neut}$, such that  
the branching ratio for $\gl \rightarrow \bone b$ decays is 100\%. Sbottoms  
are produced via gluino-mediated processes or via direct pair production. They are assumed to 
decay exclusively via $\bone \rightarrow b\neut$, where $m^{}_{\neut}$ is assumed to be 
60~GeV, above the present exclusion limit~\cite{lep2}.

In the one-lepton analysis, events are required to contain energetic 
jets, of which one must be identified as a $b$-jet, large $\etmiss$ and at least one
high-\pt\ electron or muon. This analysis is sensitive to SUSY scenarios in which 
the stop is the lightest squark and 
$m^{}_{\gl}>m^{}_{\tone}$. 
If the stop decay channel $\tone \rightarrow b \chipm$ dominates, 
possible subsequent $ \chipm \to \neut l^{\pm} \nu$ decays result in experimental 
signatures with energetic charged leptons in addition to $b$-jets and 
$\etmiss$. In the present analysis, only  
$\gl\gl$ and $\tone\tone$ pair production 
are considered, with 100\% branching ratios for the 
$\gl \rightarrow \tone t$ and $\tone \rightarrow b \chipm$ decays. The chargino is 
assumed to have a mass $m^{}_{\chipm} \simeq 2 \cdot m^{}_{\neut}$, with 
$m^{}_{\neut}=60$~GeV, and to decay through a virtual $W$ boson 
(BR($ \chipm \to \neut l^{\pm} \nu$)=11\%). 

In addition to the aforementioned phenomenological MSSM models, the results 
are interpreted in the framework of minimal supergravity (MSUGRA/CMSSM~\cite{msugra}) 
and in specific Grand Unification Theories (GUTs) based on the gauge group 
SO(10)~\cite{SO10}. For \msugra, limits on the universal scalar and gaugino mass 
parameters ($m_0,m^{}_{1/2}$) are presented for fixed values of the ratio of the 
Higgs vacuum expectation value, tan$\beta$=40, the common trilinear coupling at 
the GUT scale $A_0$=0~GeV($-500$~GeV), and the sign of the Higgsino mixing 
parameter $\mu>0$. Taking large values of tan$\beta$ or negative values of $A_0$ with 
other model parameters held fixed leads to lower third generation sparticle masses 
compared to those of the other sparticles. 
Depending on $m_0$ and $m^{}_{1/2}$, any of the final states 
such as $\sq\sq$, $\sq\gl$ and $\gl\gl$ might be dominant. 
In the SO(10) scenario, 
the SUSY particle mass spectrum is characterised by the low masses of the 
gluinos (300-600 GeV), 
charginos (100-180 GeV) and neutralinos (50-90 GeV), whereas all scalar particles have masses 
beyond the TeV scale. Depending on the sparticle masses, chargino-neutralino and
gluino-pair production dominate. The three-body gluino decays $\gl \rightarrow b\bar{b}\tilde{\chi}^{0}_{1}$ and 
$\gl \rightarrow b\bar{b}\tilde{\chi}^{0}_{2}$  are expected to lead to final states 
with high $b$-jet multiplicities. Two specific models are considered~\cite{SO10HSDR3}, 
the D-term splitting model, DR3, and the Higgs splitting model, HS.

%% file: detector.tex
The ATLAS detector~\cite{DetectorPaper:2008} comprises an inner detector surrounded by a thin superconducting solenoid, 
and a calorimeter system. Outside the calorimeters is an
extensive muon spectrometer in a toroidal magnetic field.

  The inner detector system is immersed in a $2$\,T axial magnetic field and provides 
tracking information for charged particles in a pseudorapidity 
range $|\eta|< 2.5$.\footnote{The azimuthal angle $\phi$ is measured around the beam axis and the polar 
angle $\theta$ is the angle from the beam axis. The pseudorapidity is defined as $\eta = -\ln \tan(\theta/2)$. 
The distance~$\Delta R$ in the $\eta-\phi$ space is defined as $\Delta R = \sqrt{(\Delta\eta)^2+(\Delta\phi)^2}$.} 
The highest granularity is achieved around the vertex region using silicon pixel and microstrip 
detectors. These detectors allow for an efficient tagging of jets originating from $b$-quark decays 
using impact parameter measurements and the reconstruction of secondary decay vertices.
The transition radiation tracker, which surrounds the silicon detectors, contributes to track reconstruction 
up to $|\eta| = 2.0$ and improves the electron identification by the detection of transition radiation.

  The calorimeter system covers the pseudorapidity range $|\eta| < 4.9$. 
The highly segmented electromagnetic calorimeter consists of lead absorbers with liquid argon 
as the active material and covers the pseudorapidity range $|\eta| < 3.2$. In the region  $|\eta| < 1.8$, a presampler detector consisting of a thin 
layer of liquid argon is used to correct for the energy lost by electrons, positrons, and photons upstream of the calorimeter.
The hadronic tile calorimeter is a steel/scintillating-tile detector and is placed directly 
outside the envelope of the electromagnetic calorimeter. In the forward regions, 
it is complemented by
two end-cap calorimeters using liquid argon as active material and copper or tungsten as absorber 
material. 

Muon detection is based on the magnetic deflection of muon tracks in the large superconducting air-core toroid magnets, instrumented with separate 
trigger and high-precision tracking chambers. A system of three toroids, a barrel and two end-caps, generates the magnetic field for the muon spectrometer in the pseudorapidity range $|\eta|<2.7$.

%% file: MCsample_paper.tex
Simulated event samples were used to determine the detector acceptance, the 
reconstruction efficiencies and the expected event yields for signal and 
background 
processes. 

SUSY signal processes were generated for various models using the 
{\tt HERWIG++}~\cite{herwig} v2.4.2 Monte Carlo program. 
The particle mass spectra and decay modes were determined using the 
{\tt ISASUSY} from {\tt ISAJET}~\cite{isajet} v7.80 and {\tt SUSYHIT}~\cite{susyhit} v1.3 
programs. The latter was used for the assumed MSSM scenarios, which are parametrised 
in the ($m^{}_{\gl},m^{}_{\bone}$) and ($m^{}_{\gl},m^{}_{\tone}$) planes, 
with gluino masses above 300~GeV. 
The SUSY sample yields were normalised to the results of 
next-to-leading order (NLO) calculations, as obtained using 
the {\tt PROSPINO}~\cite{prospino} v2.1 program. 
For these calculations the CTEQ6.6M~\cite{cteq6m} parametrisation of the parton density functions (PDFs)
was used and the renormalisation and factorisation scales were set to the 
average mass of the sparticles produced in the hard interaction.   

\begin{table}[t]
\begin{small}
\begin{center}
\begin{tabular}{ l|r r}
\hline
\hline
\raisebox{-0.4ex}{Physics process} &  {\raisebox{-0.4ex}{$\sigma \cdot$ BR [nb]}} &  \\
\hline
\hline
$W \rightarrow \ell \nu$ (+jets) & 31.4$\pm$1.6  &\cite{Hamberg:1990np,Melnikov:2006kv,Melnikov:2006di}  \\
$Z/\gamma^* \to \ell^+ \ell^-$ (+jets) & 3.20$\pm$0.16 &\cite{Hamberg:1990np,Melnikov:2006kv,Melnikov:2006di} \\

$Z \to \nu \bar \nu$ (+jets)  & 5.82$\pm$0.29 &\cite{Hamberg:1990np,Melnikov:2006kv,Melnikov:2006di} \\

$\ttbar$          & 0.165$^{+0.011}_{-0.016}$  &\cite{Bonciani:1998vc,Moch:2008qy,Beneke:2009ye}\\
Single top         & 0.037$\pm$0.002  &\cite{Bonciani:1998vc,Moch:2008qy,Beneke:2009ye}\\
Dijet ($\hat{p}_{\mathrm{T}}> 8$~GeV)  & 10.47$ \times 10^6$&\cite{pythia} \\
\hline
\hline
\end{tabular}
\caption{The most important background processes and their production cross sections, 
multiplied by the relevant branching ratios (BR). Contributions from higher order 
QCD corrections are included for  
$W$ and $Z$ boson production (NNLO corrections) and for \ttbar\ production 
(NLO+NNLL corrections). 
The inclusive QCD jet cross section is given at leading order (LO). 
The QCD sample was generated with a cut on the 
transverse momentum of the partons involved in the 
hard-scattering process, $\hat{p}_{\mathrm{T}}$.}
\label{MC}
\end{center}
\end{small}
\end{table}

For the backgrounds the following Standard Model processes were considered: 
\begin{itemize} 
\item $\ttbar$ and single top production: events were generated using the generator 
{\tt MC@NLO}~\cite{mcatnlo,mcatnlo2} v3.41. For the evaluation of systematic uncertainties, additional 
$\ttbar$ samples were generated using the {\tt POWHEG}~\cite{powheg} and {\tt ACERMC}~\cite{Kersevan:2004yg} 
programs. 

\item $W (\rightarrow \ell \nu)$+jet, $Z/\gamma^*(\rightarrow \ell^+ \ell^-)$+jet  
(where $\ell=e,\mu,\tau$) and 
$Z(\rightarrow \nu \bar{\nu})$ +jet 
production: events with light and heavy ($b$) flavour jets were generated using the 
{\tt ALPGEN}~\cite{alpgen} v2.13 program. A generator level cut $m_{\ell 
\ell}>40$~GeV was 
applied to the $Z/\gamma^*(\rightarrow \ell^+ \ell^-)$ process.

\item Jet production via QCD processes (referred to as ``QCD background'' in the following):  
events were generated using the {\tt PYTHIA}~\cite{pythia} v6.4.21 generator. 
For the evaluation of systematic uncertainties, samples produced with {\tt ALPGEN} were used.

\item Di-boson ($WW$, $WZ$ and $ZZ$) production: events were generated using {\tt ALPGEN}, 
however, compared to the other backgrounds their
contribution was found to be negligible, after the application of the selection criteria. 
\end {itemize} 

All signal and background samples were generated at $\sqrt{s}=7$~TeV using the ATLAS 
MC09 parameter tune~\cite{mc09tunes}, processed with the {\tt GEANT4}~\cite{geant4}~simulation 
of the ATLAS detector~\cite{atlassimulation}, and then reconstructed 
and passed through the same analysis chain as the data. For all generators, except for 
{\tt PYTHIA}, the {\tt HERWIG + JIMMY}~\cite{herwig, Jimmy} modelling of the parton shower and 
underlying event was used (v6.510 and v4.31, respectively). 

For the comparison to data, all background cross sections, except the QCD background cross section, 
were normalised to the results of higher order QCD calculations. 
A summary of the relevant cross sections is given in Table~\ref{MC}.
For the next-to-next-to-leading order (NNLO) $W$ and $Z/\gamma^*$ production cross sections, an uncertainty of $\pm$5\% is 
assumed \cite{w-paper}. For the $\ttbar$ production cross section, the 
corresponding uncertainty on the NLO+NNLL (next-to-next-to-leading logarithms) cross section was estimated to be $^{+6.5\%}_{-9.5\%}$.
For the QCD background, no reliable prediction can be obtained from a leading order 
Monte Carlo simulation and data-driven methods were used to determine the residual contributions 
of this background to the selected event samples, as discussed in Section~\ref{sec:wbkg}.

%% file: data_evtsel.tex
After the application of beam, detector and 
data-quality requirements, the data set used for this analysis resulted 
in a total integrated luminosity of 35 pb$^{-1}$. 

For the zero-lepton analysis, events were selected at the trigger level by 
requiring jets with high transverse momentum. The selection is fully efficient 
for events containing at least one jet with $\pt >120$~GeV. A further trigger 
level requirement of $\etmiss>25$~GeV was applied~\cite{MissingEtTrigger}. 
For the one-lepton analysis, the trigger selection was based on single lepton
triggers, which retain events if an electron with $\pt > 15$~GeV  
or a muon with $\pt > 13$~GeV is present within the trigger acceptance.  

In the data sample selected, jet candidates were reconstructed by using the anti-$k_t$ jet 
clustering algorithm~\cite{anti-kt, Cacciari200657, Fastjet} with a distance parameter of $R$=0.4. The inputs to this 
algorithm are three dimensional topological calorimeter energy clusters. 
The jet energies were corrected for inhomogeneities and for the non-compensating nature 
of the calorimeter by using $\pt$- and $\eta$-dependent 
calibration factors. They were determined from Monte Carlo simulation and 
validated using extensive test-beam measurements and studies of $pp$ collision data (Ref.~\cite{JES} and references therein). 
Only jets with $\pt>20$~GeV and within $|\eta|<2.5$ were retained. 
Candidates for $b$-jets were identified among jets with $\pt>30$~GeV using 
an algorithm that reconstructs a  
vertex from all tracks which are displaced from the primary vertex and associated with the jet. 
The parameters of the algorithm were chosen such that a tagging efficiency of 50\% (1\%) was 
achieved for $b$-jets (light flavour or gluon jets) in $\ttbar$ events in Monte Carlo simulation~\cite{toppaper}.

Electron candidates were required to 
satisfy the `medium' (zero-lepton analysis) or `tight' (one-lepton analysis) selection criteria, as detailed in  
Ref.~\cite{elrecon}. 
Muon candidates were identified either as a match between an extrapolated inner 
detector track and one or more segments in the muon spectrometer, or by associating 
an inner detector track to a muon spectrometer track. The combined track parameters 
were derived from a statistical combination of the two sets of track parameters.
Electrons and muons were required to have $\pt>20$ GeV and $|\eta|<2.47$ 
or $|\eta|<2.4$, respectively.

The calculation of $\etmiss$ is based on the
modulus of the vectorial sum of the $\pt$ of the reconstructed jets
(with $\pt > 20 \GeV$ and over the full calorimeter coverage $|\eta| < 4.9$), leptons 
  (including non--isolated muons) and the
calorimeter clusters not belonging to reconstructed objects.

After object identification, overlaps were resolved.
Any jet within a distance $\Delta R $ = 0.2
of a medium electron candidate was discarded. 
The whole event was rejected if one or more electrons were identified in the transition 
region $1.37 < |\eta|<1.52$ between the barrel and endcap calorimeters. 
Any remaining lepton within $\Delta R=$ 0.4 of a jet was discarded. 

Events were selected if a reconstructed primary vertex was found associated with 
five or more tracks, and if they passed basic quality criteria against detector noise and 
non-collision backgrounds. 

In the zero-lepton analysis, events were required to have at least one jet with 
$\pt>120$~GeV, two additional jets with $\pt>30$~GeV and $\etmiss>100$~GeV. At least one jet is required to be $b$-tagged. 
Events containing identified  `medium' electron or muon candidates were rejected.   
The effective mass, $\meff$, is defined as the scalar sum of $\etmiss$ and the 
transverse momenta of the highest $\pt$ jets (up to a maximum of four). Events were required to 
have $\etmiss / \meff>0.2$. In addition, the smallest azimuthal separation between the 
$\etmiss$ direction and the three leading jets, $\dphimin$, was required to be larger 
than 0.4. 
The last requirement reduces the amount of QCD  background effectively since, in 
this case, $\etmiss$ results from mis-reconstructed jets or from 
neutrinos emitted along the direction of the jet axis by heavy flavour decays.

In the one-lepton analysis, events were required to have at least one muon or 
a `tight' electron, two jets with $\pt>60$~GeV and 
$\pt>30$~GeV respectively, $\etmiss>80$~GeV and $\trmass>100$~GeV, 
where $\trmass$ is the transverse mass constructed using the highest $\pt$ lepton and 
$\etmiss$. At least one jet is required to be $b$-tagged.  The $\trmass$ cut rejects events with a $W$ boson in the final state.

In both analyses, further cuts on $\meff$ were applied 
to maximise the sensitivity to gluino-mediated production of 
sbottoms or stops.  
A threshold on \meff at 600 GeV (500 GeV) was chosen for 
the zero-lepton (one-lepton) analysis. It should be noted that for the one-lepton analysis  
the transverse momenta of reconstructed leptons are included in the definition of the 
$\meff$.

The event selection efficiency for each SUSY signal hypothesis was calculated as the sum of 
the efficiencies for the $\gl\gl$ and $\bone\bone$ ($\tone\tone$) processes, weighted by 
their respective NLO cross sections. 
For the zero-lepton selection, the efficiency varies between 7\% and 50\% across the 
($m^{}_{\gl}$,$m^{}_{\bone}$) plane. The lowest values are found at large 
$\Delta{m}$= $m^{}_{\gl}-m^{}_{\bone}$, where the production of 
$\bone\bone$ pairs dominates. As $\Delta{m}$ decreases, high efficiency values are 
found down to $\Delta{m} \simeq 20$~GeV. For the one-lepton channel, 
the efficiency for ($\gl,\tone$)-type SUSY signals  varies between 0.4\% and 3\% across 
the ($m^{}_{\gl}$,$m^{}_{\tone}$) plane and depends on $\Delta{m}$~=~$m^{}_{\gl}$-$m^{}_{\tone}$ 
in a similar way to the gluino-sbottom case. 

No additional dedicated optimisations were performed for the \msugra and SO(10) scenarios.  
The efficiencies for the zero-lepton (one-lepton) selection for \msugra range between 
8\% (1\%) for $m^{}_{1/2} \simeq$~130~GeV and 23\% (12\%) for $m^{}_{1/2} \simeq$~340~GeV, 
with a smaller dependence on $m_0$. For SO(10) models, the highest sensitivity
is reached in the zero-lepton analysis, with dominant contributions via  
$\gl\gl$ production. In this case, the efficiencies vary between 7\% and 20\% as the 
gluino mass increases and are generally found to be larger for the DR3 scenario than for
 the HS scenario. 

%% file: bkg_paper.tex
Standard Model processes contribute to the events that survive 
the selection described in the previous section. The dominant source is 
\ttbar\ production due to the presence of jets, $\etmiss$ and $b$-quarks in the final state.

The QCD background to the zero-lepton final state was estimated 
by normalising the PYTHIA Monte 
Carlo prediction to data in a QCD-enriched control region defined by $\dphimin<0.4$.  
The Monte Carlo was then used to evaluate the ratio between the number of 
events in this control region and the signal region ($\dphimin>0.4$).
In the one-lepton final state the number of QCD multi-jet events was estimated 
using a matrix method similar to the one described in Ref.~\cite{w-paper}. 
Cuts on the electron and muon identification were relaxed 
to obtain ``loose'' control samples that are dominated by QCD jets. 

The non-QCD background in the zero-lepton final state was estimated using 
Monte Carlo simulation, while in the case of the one-lepton final state a data-driven technique 
is employed. This method exploits the low correlation between
 $\meff$ and $\trmass$. 
Four regions were defined: 
(A) $40<\trmass<100$~GeV and $\meff<500$~GeV, 
(B) $\trmass>100$~GeV and $\meff<500$~GeV, 
(C) $40<\trmass<100$~GeV and $\meff>500$~GeV and 
(D) $\trmass>100$~GeV and $\meff>500$~GeV. 
Regions A-C are dominated by background from \ttbar\ and $W$+jet production. 
Assuming that the variables are uncorrelated, the number of background events in the signal 
region is given by $N_D=N_C \times N_B/N_A$, where $N_A$, $N_B$, $N_C$ are the numbers of events in the regions A, B and C, respectively.
A Monte Carlo simulation was used to validate the method and to 
establish possible sources of systematic 
uncertainties. The small number of events in the control regions is the main 
limitation of the method.
It was also checked that a SUSY signal 
contamination does not bias the estimated background  
 and that any bias is smaller than the systematic uncertainties assigned to the method and on the expected SUSY prediction.

%% file: syst_paper.tex
Various systematic uncertainties affecting signal and background rates 
were considered.

For the zero-lepton analysis, the backgrounds from \ttbar\ and $W/Z$+jet production are
taken from Monte Carlo simulation. The total uncertainty on this prediction was estimated to 
be $\pm$35\% after the final selection. It is dominated by the uncertainty on the jet energy 
scale (JES)~\cite{JES}, the uncertainty on the theoretical prediction of the 
background processes and the uncertainty on the determination of 
the \btagging efficiency~\cite{toppaper}. 
The uncertainty on the jet energy scale varies as a function of jet \pT, and 
decreases from 6\% at 20~\GeV\ to 4\% at 100~\GeV, with additional contributions 
taking into account the dependence of the jet response on the jet isolation and flavour. 
This translates into a $\pm$25\% uncertainty on the absolute prediction of the 
background from SM processes.  
Uncertainties on the theoretical cross sections of the background processes (see Section~\ref{sec:SimEvSamp}), 
on the modelling of initial and final-state soft gluon radiation and the limited knowledge of the 
PDFs of the proton 
lead to uncertainties of $\pm$20\% and $\pm$25\% on the absolute predictions of the \ttbar\ and  
the $W/Z$+jet backgrounds, respectively.  
The uncertainty on the determination of the tagging 
efficiency for \bjets, \cjets and \ljets introduces further 
uncertainties on the predicted background contributions at the level of 
$\pm$12\% for \ttbar\ and 
$\pm$25\% for $W/Z$+jets. 
Other uncertainties result from the modelling of additional 
pile--up interactions ($\pm$5\%) and 
of the trigger efficiency ($\pm$3\%) in the Monte Carlo simulation. 
For the QCD  background estimation, the uncertainty is dominated by the limited number of Monte 
Carlo events available for the zero-lepton analysis.

For the one-lepton analysis, where a data-driven technique was employed, 
the small event number in the control regions was the dominant uncertainty 
($\pm$25\%). 
Residual uncertainties associated to the 
method due to the JES, \btagging, lepton identification and theoretical predictions of the relative 
contributions of $W$ and \ttbar\ backgrounds  
were studied using Monte Carlo simulation and estimated to be at the level 
of $\pm$8\%.

For the SUSY signal processes, various sources of uncertainties 
affect the theoretical NLO cross sections.
Variations of the renormalisation and factorisation scales by a factor 
of two result in uncertainties of $\pm$16\% for 
$\gl\gl$ production and $\pm$30\% ($\pm$27\%) for $\bone\bone$ ($\tone\tone$) pair 
production, almost independently of the sparticle mass and the 
SUSY model. Uncertainties for $\sq\sq$ and $\sq\gl$ production, relevant in \msugra scenarios,    
were estimated to be at the level of $\pm$10\% and $\pm$15\%, respectively. 

The number of predicted signal events is also affected by the 
PDF uncertainties, estimated using the CTEQ 6.6M PDF error 
eigenvector sets at the 90\% C.L. limit, and rescaled by 1/1.645. 
The relative uncertainties on the $\gl\gl$ ($\bone\bone$, $\tone\tone$) cross sections
were estimated to be in the range from $\pm$11\% to $\pm 25\%$ 
($\pm 7\%$ to $\pm 16\%$) for the $\gl\gl$ ($\bone\bone$, $\tone\tone$) processes, depending on 
the gluino (sbottom, stop) masses.
For first and second generation squark-pair and associated gluino-squark 
 production, the uncertainty on the PDFs varies between 
$\pm 5\%$ and $\pm 15\%$ as the squark masses increase.  
The impact of detector related uncertainties, such as the JES and \btagging,
on the signal event yields depends on the masses of the 
most copiously produced sparticles. The total uncertainty varies between 
$\pm$25\%  and $\pm$10\% as the gluino/squark masses increase from 300~GeV 
to 1~TeV, across the different scenarios, and it is dominated by the JES 
and the \btagging uncertainty for low and high mass sparticles, respectively. 

Finally, an additional $\pm$11\% uncertainty on the quoted total integrated 
luminosity was taken into account~\cite{Aad:2011dr}.

%% file: results.tex
\begin{figure*}[t]
  \begin{minipage}{0.49\linewidth}
  \begin{center}
    \includegraphics[width=1.\linewidth]{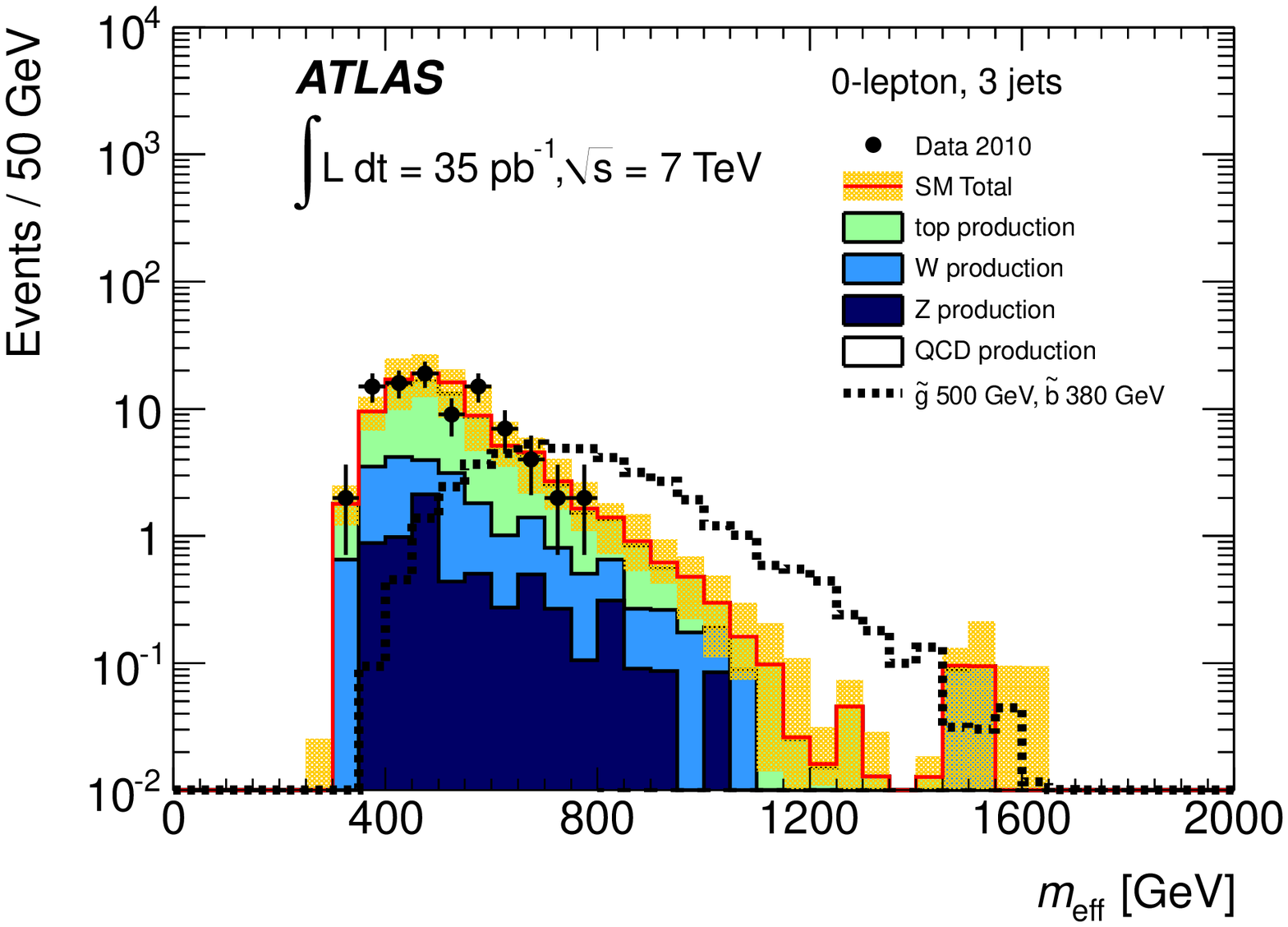}
    \includegraphics[width=1.\linewidth]{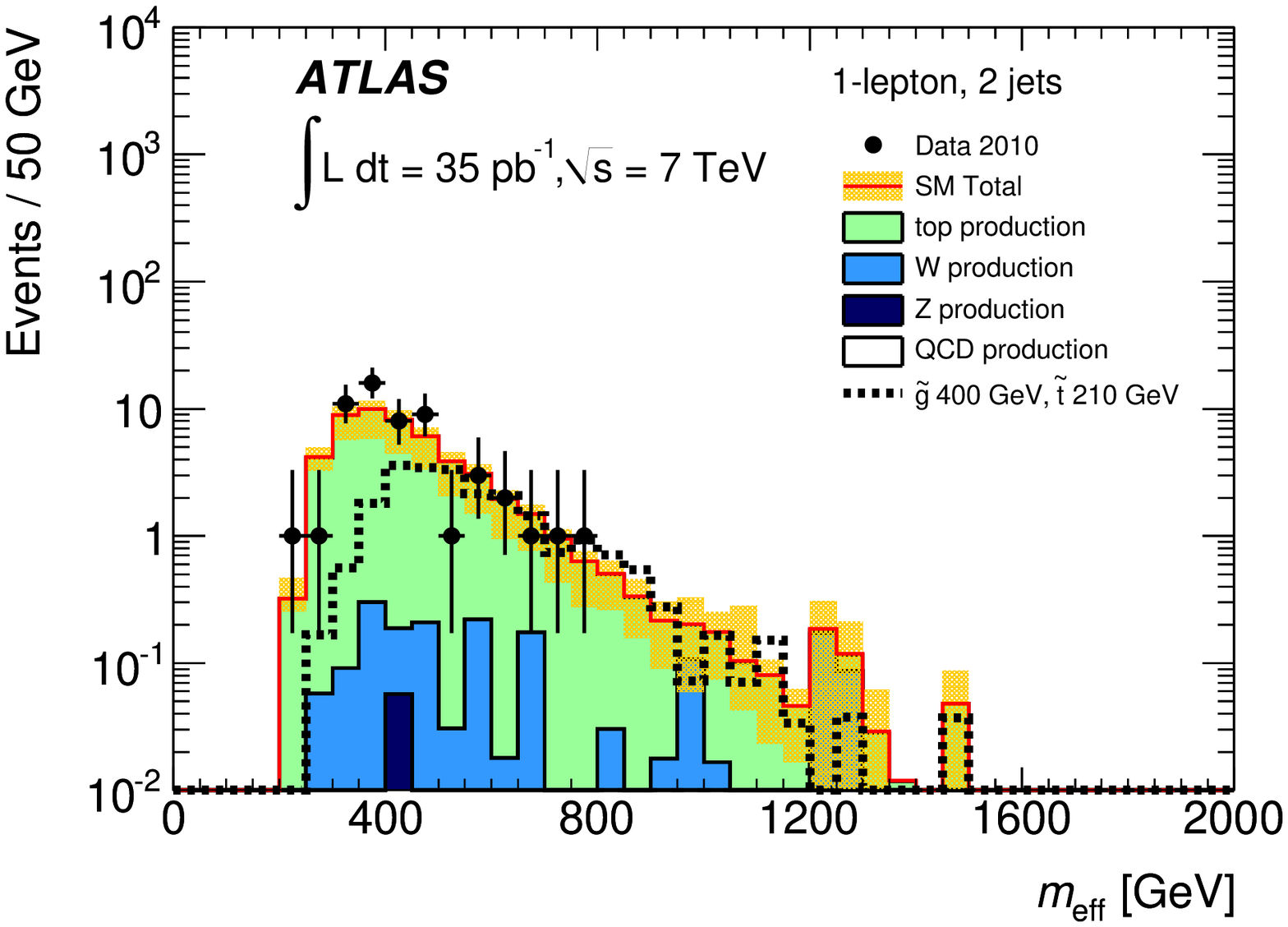}
  \end{center}
  \end{minipage}
  \begin{minipage}{0.49\linewidth}
  \begin{center}
    \includegraphics[width=1.\linewidth]{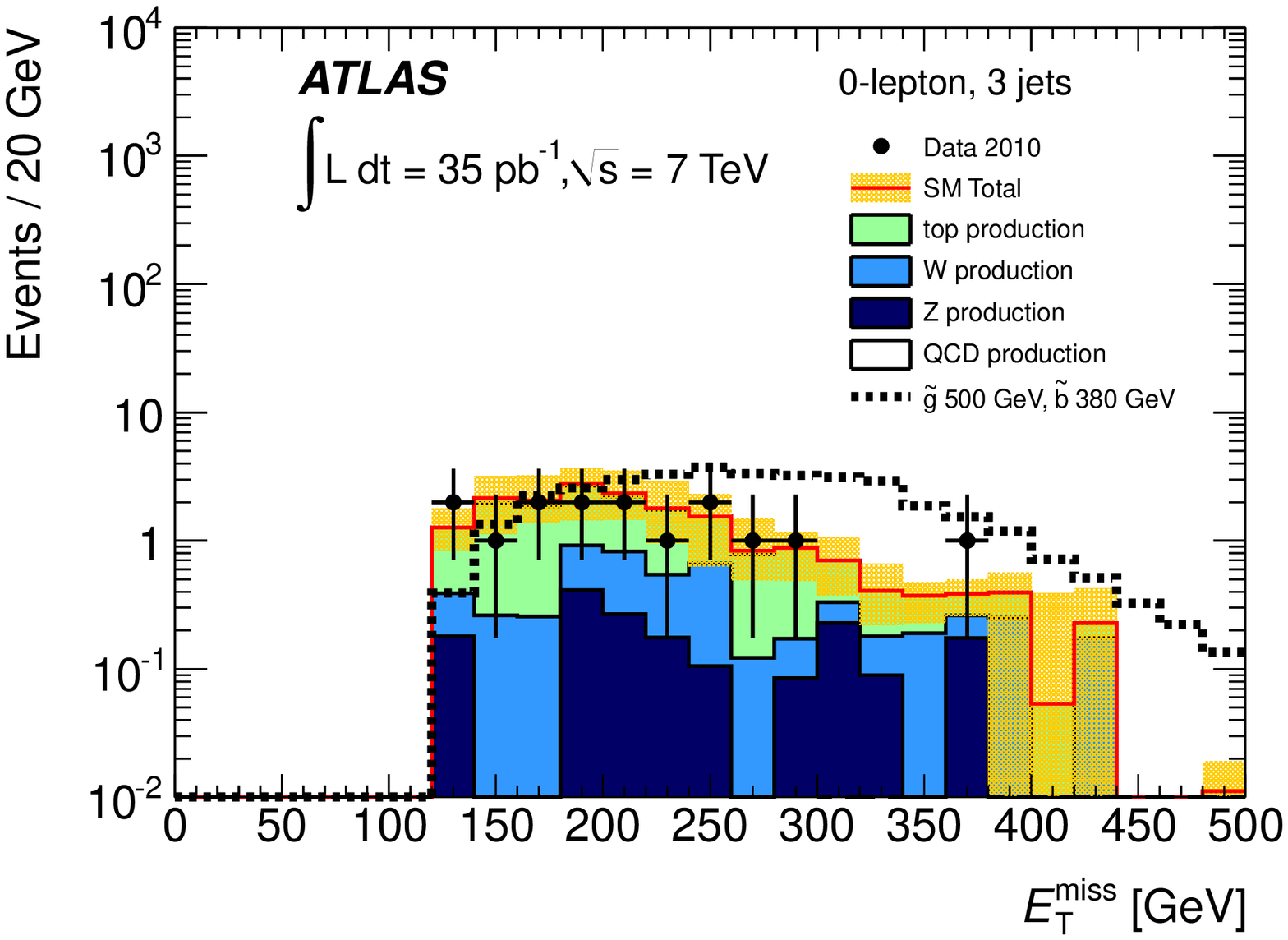}
    \includegraphics[width=1.\linewidth]{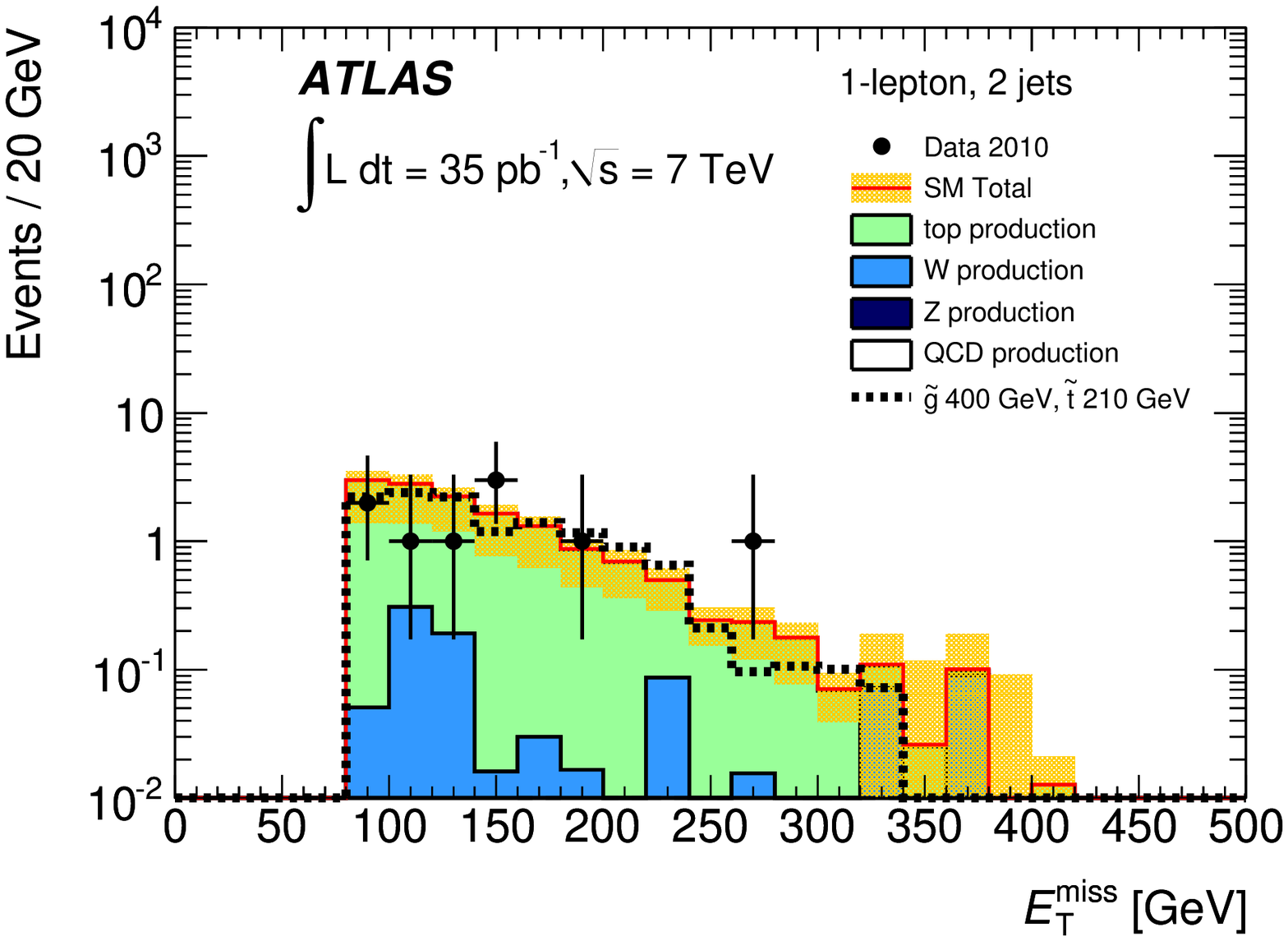}
  \end{center}
  \end{minipage}
  \caption{Distributions of the effective mass, $\meff$, (left) and the $\met$, (right) for data 
and for the expectations from Standard Model processes after the baseline selections in the 
zero-lepton(top) and one-lepton channel (bottom). 
The data correspond to an integrated luminosity of 35~\ipb. 
Black vertical bars show the statistical uncertainty of the data. The yellow band shows 
the full systematic uncertainty on the SM expectation. 
The $\met$ distributions are shown after a cut on $\meff$ at 600 GeV (zero-lepton) and 500 GeV (one-lepton). 
For illustration, the distributions for one reference SUSY signal, relevant for each channel, 
are superimposed. 
}
  \label{fig:SignalKin}
\end{figure*}
In Figure~\ref{fig:SignalKin} the distributions of \meff and of \etmiss 
are shown for the two analyses. For the \etmiss\ distributions all 
cuts described in Section 4 are applied. The \meff\ distributions are shown 
after the application of all cuts, except for the \meff\ cut. 

The expectations from Standard Model background
processes are superimposed. 
For illustration, the figures also include the distributions expected for 
SUSY signals. For the zero-lepton channel, a scenario with $m^{}_{\gl}=500$ GeV 
and $m^{}_{\bone}=380$ GeV is chosen, while for the one-lepton channel the relevant masses 
are $m^{}_{\gl}=400$ GeV 
and $m^{}_{\tone}=210$~GeV.
In Table~\ref{tab:summary}, 
the observed number of events and the predictions for contributions from 
Standard Model processes are presented. For both analyses, the 
data are in agreement with the Standard Model predictions, within uncertainties.

\begin{table}[btc]
\begin{small}
\begin{center}
\begin{tabular}{lccc}
\hline
\hline
 & 0-lepton &1-lepton & 1-lepton   \\
 &          & Monte Carlo & data-driven \\ 
\hline
\hline
\ttbar\ and single top & $12.2 \pm 5.0$    & $12.3 \pm 4.0$&  $14.7 \pm 
3.7$\\
W and Z & $6.0 \pm 2.0$ & $0.8 \pm 0.4 $& - \\
QCD & $1.4 \pm 1.0$  & $0.4\pm 0.4$ & $0^{+0.4}_{-0.0}$\\
\hline
\hline
Total SM & $19.6 \pm 6.9$ & $13.5 \pm 4.1$ & $14.7 \pm 3.7$ \\
Data & 15 & 9 & 9 \\
\hline
\hline
\end{tabular}
\caption{Summary of the expected and observed event yields. The QCD prediction for the 
zero-lepton channel is based on the semi-data-driven method 
described in the text. For the one-lepton channel, the results for both the Monte Carlo 
and the data-driven approach are given. Since the data-driven technique does not distinguish 
between top and W/Z backgrounds the total background estimate is shown in the 
top row. The errors are systematic for the expected Monte Carlo prediction and statistical for the data-driven technique.} \label{tab:summary}
\end{center}
\end{small}
\end{table}

The results are translated into 95\% C.L. upper limits on 
contributions from new physics. Limits were derived using a profile likelihood ratio,  
$\Lambda(s)$, 
where the likelihood function of the fit was written as 
$L({n}|s,{b},{\btheta}) = P_{\mathrm{S}} \times \, C_{\mathrm{Syst}}$;      
${n}$ represents the number of observed events in data, $s$ is the SUSY signal under 
consideration, ${b}$ is the background, and ${\btheta}$ represents the systematic 
uncertainties. The $P$ function is a Poisson-probability distribution for event counts 
in the defined signal region and $C_{\mathrm{Syst}}$ represents the constraints on 
systematic uncertainties, which are treated as nuisance parameters with a Gaussian 
probability density function and correlated when appropriate. 
The exclusion $p$-values are obtained from the test statistic $\Lambda(s)$ using 
pseudo-experiments and one-sided upper limits are set~\cite{higgscombi}.
	
Upper limits at 95\% C.L. on the number of signal events in the signal   
regions are obtained independently of new physics models for the zero- and one-lepton 
final states. These numbers are 10.4 and 4.7, respectively, and correspond to 
95\% C.L. upper limits on effective cross sections for new processes of 
0.32~pb and 0.13~pb for the zero- and one-lepton channel, respectively.   
The cross section upper limits include the $\pm 11\%$ uncertainty on the quoted 
total integrated luminosity. 

\begin{figure}[t]
  \begin{center}
    \includegraphics[width=1.\linewidth]{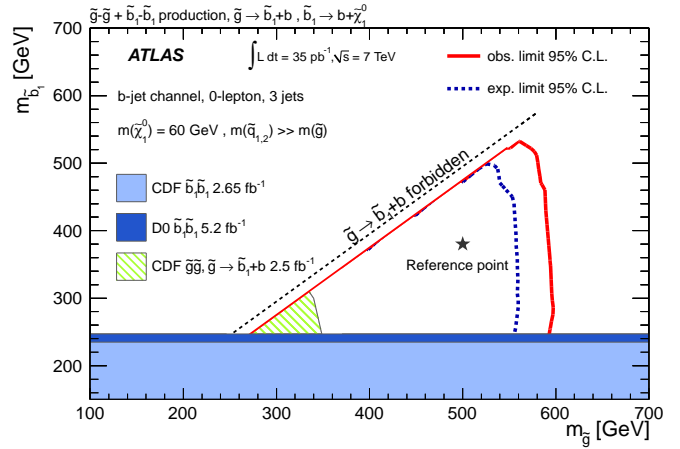}
  \end{center}
  \caption{Observed and expected 95\% C.L. exclusion limits, as obtained with the zero-lepton channel, 
  in the ($m^{}_{\gl},m^{}_{\bone}$) plane. The neutralino mass is assumed to be 60~GeV and the NLO cross sections are 
  calculated using {\tt PROSPINO} in the hypothesis of $m^{}_{\sqonetwo} \gg m^{}_{\gl}$. 
  The result is compared to previous results from CDF searches which assume the same gluino-sbottom 
  decays hypotheses, a neutralino mass of 60~GeV and $m^{}_{\sqonetwo}=500$~GeV ($\gg m^{}_{\gl}$ for the Tevatron 
  kinematic range). Exclusion limits from the CDF and D0 experiments on direct sbottom pair 
  production~\cite{sbottomCDF, sbottomD0} are also reported. }
  \label{fig:sbottomgluino}
\end{figure}

These results can be interpreted in terms of 95\% C.L. 
exclusion limits in several SUSY scenarios. 
In Figure~\ref{fig:sbottomgluino} the observed and expected exclusion regions are shown
in the ($m^{}_{\gl},m^{}_{\bone}$) plane, for the hypothesis that the lightest 
squark $\bone$ is produced via gluino-mediated or direct pair production and 
decays exclusively via $\bone \rightarrow b\neut$. The zero-lepton channel was  
considered for this model and the largest acceptance was found for $\gl\gl$ 
production. The limits do not strongly depend on the neutralino mass assumption 
as long as $m^{}_{\gl}-m^{}_{\neut}$ is larger than 250-300~GeV, due to the 
harsh kinematic cuts. Gluino masses below 590~GeV are excluded for 
sbottom masses up to 500~GeV. These limits depend weakly -- via the 
dependence of the production cross section for $\gl\gl$ production --  
on the masses of the first and second generation squarks, $\sqonetwo$. 
Variations of these masses in the range between $\sim$3~TeV  
and $2 \cdot m^{}_{\gl}$ reduce the excluded mass region by less than 20 GeV.  

The zero-lepton analysis was also employed to extract limits on the gluino 
mass in the two SO(10) scenarios, DR3 and HS. 
Gluino masses below 500 GeV are excluded for the DR3 models considered, 
where $\gl \rightarrow b\bar{b}\tilde{\chi}^{0}_{1}$ decays dominate.  
A lower sensitivity ($m^{}_{\gl}<420$~GeV) was found for the HS model, 
where larger branching ratios of $\gl \rightarrow b\bar{b}\tilde{\chi}^{0}_{2}$ 
are expected and the efficiency of the selection is reduced with respect to 
the DR3 case. 

\begin{figure}[t!]
    \includegraphics[width=0.875\linewidth]{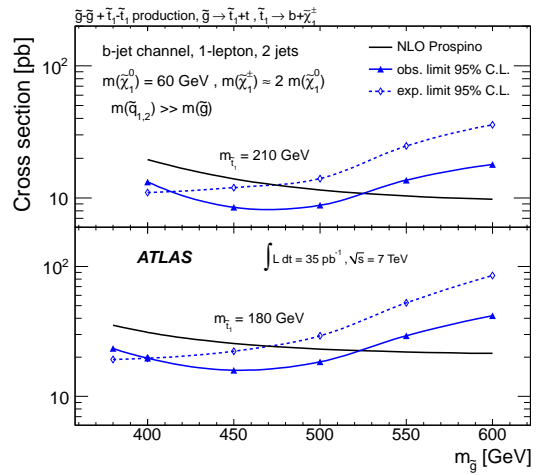}
  \caption{Observed and expected 95\% C.L. upper limits, as obtained with the one-lepton analysis, on 
  the gluino-mediated and stop pair production cross section as a function of the gluino mass for 
  two assumed values of the stop mass and BR($\tone \rightarrow b\chipm)=1$. The chargino is 
  assumed to have twice the mass of the neutralino ($=60$~GeV) and NLO cross sections are calculated 
  using {\tt PROSPINO} in the hypothesis of $m^{}_{\sqonetwo} \gg m^{}_{\gl}$.   
  Theoretical uncertainties on the NLO cross sections are included in the limit calculation. }
  \label{fig:stopgluino}
\end{figure}

The results of the one-lepton analysis were interpreted as exclusion limits on the 
($m^{}_{\gl},m^{}_{\tone}$) plane in the hypothesis that the lightest $\tone$ is 
produced via gluino-mediated or direct pair production. Stop masses above 130~GeV 
are considered, and $\tone$ is assumed to decay exclusively via $\tone \rightarrow b \chipm$.  
In Figure~\ref{fig:stopgluino} the observed and expected exclusion limits 
are shown as a function of $m^{}_{\gl}$ for two representative values of the 
stop mass. Gluino masses below 520~GeV are excluded for stop masses in 
the range between 130 and 300~GeV. 

\begin{figure}[t]
  \begin{center}
    \includegraphics[width=1.\linewidth]{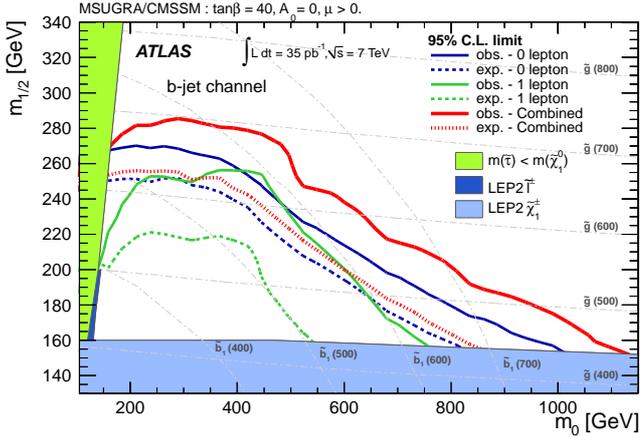}
  \end{center}
  \caption{Observed and expected 95\% C.L. exclusion limits as obtained from the zero- and one-lepton 
  analyses, separately and combined, on \msugra scenario with tan$\beta$ = 40, $A_0$ = 0, $\mu>0$. The light-grey 
  dashed lines are the iso-mass curves for gluinos and sbottom -- stop masses are 15\% lower than sbottom 
  masses, across the ($m_0,m^{}_{1/2}$) parameter space.   
  The results are compared to previous limits from the LEP experiments~\cite{lep2}. }
  \label{fig:msugra}
\end{figure}

Finally, the results of both analyses were used to calculate 95\% C.L. exclusion limits 
in the \msugra framework with large tan$\beta$. Figure~\ref{fig:msugra} shows the 
observed and expected limits in the ($m_0,m^{}_{1/2}$) plane, assuming tan$\beta$ = 40, and fixing 
$\mu>$0 and $A_0$ = 0. The largest sensitivity is found for the zero-lepton analysis. 
The combination of the two analyses, which  takes 
account  of correlations between systematic uncertainties of the two channels, 
is also shown. 
Sbottom and stop masses below 550~GeV and 470~GeV are excluded across 
the plane, respectively. Due to the \msugra constraints, this interpretation is also sensitive to 
first and second generation squarks. From the present analysis, masses of these 
squarks below 600~GeV are excluded for $m^{}_{\gl}\simeq m^{}_{\sq}$. 
Gluino masses below 500~GeV are excluded for the $m_0$ range between 
100~GeV and 1~TeV, independently on the squark masses. 
Changing the $A_0$ value from 0 to $-500$~GeV lead to significant variations in third generation 
squark mixing. Across the ($m_0,m^{}_{1/2}$) parameter space, sbottom and stop masses 
decrease by about 10\% and 30\%, respectively, if $A_0$ is changed from 0 to $-500$~GeV. 
The exclusion region of the one-lepton analysis, mostly sensitive to stop 
final states, extends the zero-lepton reach by about 20 GeV in 
$m^{}_{1/2}$ for $m_0<$600~GeV.

%% file: conclusion.tex
The ATLAS collaboration has presented a first search for supersymmetry in final states 
with missing transverse momentum and  at least one $b$-jet candidate
in proton-proton collisions at 7~TeV. The results are based on data 
corresponding to an integrated 
luminosity of 35~\ipb\ collected during 2010. 
These searches are sensitive to the gluino-mediated and direct production of 
sbottoms and stops, the supersymmetric partners of the third generation quarks, 
which, due to mixing effects, might be the lightest squarks. 

Since no excess above the expectations from Standard Model processes was found, the 
results are used to exclude parameter regions in various $R$-parity conserving SUSY models. 
Under the assumption that the lightest squark $\bone$ is produced via 
gluino-mediated processes or direct pair production and decays exclusively via 
$\bone \rightarrow b\neut$, gluino masses below 590~GeV are excluded with 95\% C.L. 
up to sbottom masses of 500~GeV. Alternatively, assuming that $\tone$ 
is the lightest 
squark and the gluino decays exclusively via $\gl \rightarrow \tone t$, 
and $\tone \rightarrow b \chipm$, gluino masses below 520~GeV are excluded for stop masses 
in the range between 130 and 300~GeV. 

In specific models based on the gauge group SO(10), gluinos with masses below 500~GeV 
and 420~GeV are excluded for the DR3 and HS models, respectively.

In an \msugra framework with large  $\tan \beta$, a significant region in the  
($m_0,m^{}_{1/2}$) plane can be excluded. For the parameters $\tan \beta$ = 40, 
$A_0 = 0$ and $\mu > 0 $, sbottom masses below 550~GeV and stop masses below 470~GeV 
are excluded with 95\% C.L. Gluino masses below 500~GeV are excluded for 
the $m_0$ range between 100~GeV and 1~TeV, independently on the squark masses. 

These analyses improve significantly on the regions of SUSY parameter space 
excluded by previous experiments that searched for similar scenarios.

%% file: Acknowledgement-09Mar11-collisions.tex
\section{Acknowledgements}

We thank CERN for the very successful operation of the LHC, as well as the
support staff from our institutions without whom ATLAS could not be
operated efficiently.

We acknowledge the support of ANPCyT, Argentina; YerPhI, Armenia; ARC,
Australia; BMWF, Austria; ANAS, Azerbaijan; SSTC, Belarus; CNPq and FAPESP,
Brazil; NSERC, NRC and CFI, Canada; CERN; CONICYT, Chile; CAS, MOST and
NSFC, China; COLCIENCIAS, Colombia; MSMT CR, MPO CR and VSC CR, Czech
Republic; DNRF, DNSRC and Lundbeck Foundation, Denmark; ARTEMIS, European
Union; IN2P3-CNRS, CEA-DSM/IRFU, France; GNAS, Georgia; BMBF, DFG, HGF, MPG
and AvH Foundation, Germany; GSRT, Greece; ISF, MINERVA, GIF, DIP and
Benoziyo Center, Israel; INFN, Italy; MEXT and JSPS, Japan; CNRST, Morocco;
FOM and NWO, Netherlands; RCN, Norway; MNiSW, Poland; GRICES and FCT,
Portugal; MERYS (MECTS), Romania; MES of Russia and ROSATOM, Russian
Federation; JINR; MSTD, Serbia; MSSR, Slovakia; ARRS and MVZT, Slovenia;
DST/NRF, South Africa; MICINN, Spain; SRC and Wallenberg Foundation,
Sweden; SER, SNSF and Cantons of Bern and Geneva, Switzerland; NSC, Taiwan;
TAEK, Turkey; STFC, the Royal Society and Leverhulme Trust, United Kingdom;
DOE and NSF, United States of America.

The crucial computing support from all WLCG partners is acknowledged
gratefully, in particular from CERN and the ATLAS Tier-1 facilities at
TRIUMF (Canada), NDGF (Denmark, Norway, Sweden), CC-IN2P3 (France),
KIT/GridKA (Germany), INFN-CNAF (Italy), NL-T1 (Netherlands), PIC (Spain),
ASGC (Taiwan), RAL (UK) and BNL (USA) and in the Tier-2 facilities
worldwide.

%% file: atlas_authlist.tex

\bigskip

G.~Aad$^{\rm 48}$,
B.~Abbott$^{\rm 111}$,
J.~Abdallah$^{\rm 11}$,
A.A.~Abdelalim$^{\rm 49}$,
A.~Abdesselam$^{\rm 118}$,
O.~Abdinov$^{\rm 10}$,
B.~Abi$^{\rm 112}$,
M.~Abolins$^{\rm 88}$,
H.~Abramowicz$^{\rm 153}$,
H.~Abreu$^{\rm 115}$
% ...
\bigskip

% ...
$^{10}$ Institute of Physics, Azerbaijan Academy of Sciences, Baku, Azerbaijan\\
$^{11}$ Institut de F\'isica d'Altes Energies and Universitat Aut\`onoma  de Barcelona and ICREA, Barcelona, Spain\\
% ...
$^{48}$ Fakult\"{a}t f\"{u}r Mathematik und Physik, Albert-Ludwigs-Universit\"{a}t, Freiburg i.Br., Germany\\
$^{49}$ Section de Physique, Universit\'e de Gen\`eve, Geneva, Switzerland\\
% ...
$^{88}$ Department of Physics and Astronomy, Michigan State University, East Lansing MI, United States of America\\
% ...
$^{111}$ Homer L. Dodge Department of Physics and Astronomy, University of Oklahoma, Norman OK, United States of America\\
$^{112}$ Department of Physics, Oklahoma State University, Stillwater OK, United States of America\\
% ...
$^{115}$ LAL, Univ. Paris-Sud and CNRS/IN2P3, Orsay, France\\
% ...
$^{118}$ Department of Physics, Oxford University, Oxford, United Kingdom\\
% ...
$^{153}$ Raymond and Beverly Sackler School of Physics and Astronomy, Tel Aviv University, Tel Aviv, Israel\\
% ...
\end{flushleft}